\documentclass[preprint,aps,psfig,amssymb,superscriptaddress]{revtex4}
\usepackage{times,floatflt}
\usepackage{amsmath}
\usepackage[]{graphicx}

\begin{document}
\title{Ground and excited states of the bipolaron in two and three dimensions}


\author{Y.H.~Ruan}
\affiliation{Department of Physics, Zhejiang University, Hangzhou
310027, P. R. China}
\affiliation{Christian-Albrechts-Universit\"at zu Kiel, Institut
f\"ur Theoretische Physik und Astrophysik, Leibnizstrasse 15,
24098 Kiel, Germany}
\author{Q.H.~Chen}
\affiliation{Department of Physics, Zhejiang University, Hangzhou
310027, P. R. China}
\author{M.~Bonitz}
\affiliation{Christian-Albrechts-Universit\"at zu Kiel, Institut
f\"ur Theoretische Physik und Astrophysik, Leibnizstrasse 15,
24098 Kiel, Germany}

\begin{abstract}
The properties of large bipolarons in two and three dimensions are
investigated by averaging over the relative wavefunction of the two
electrons and using the Lee-Low-Pines-Huybrechts variational method. We
obtain the ground-state (GS) and excited-state energies of the Fr\"{o}hlich
bipolaron for the whole range of electron-phonon coupling constants.
Furthermore, we calculate the energies of the first relaxed excited state
(RES) and Franck-Condon (FC) excited state of the bipolaron. Compared with
the FC state, the first RES has a lower energy. Our results for the GS and
RES energies are lower than those obtained before by the Landau-Pekar method
in the whole coupling regime.
\end{abstract}

\maketitle                   

\section{Introduction}\label{intro}

Polaronic and bipolaronic effects are found in many materials such
as transition metal oxides \cite{AM}, polymers \cite{DHPARK}, and
superconducting materials \cite{YHK}. Since these materials possess
low dimensional structures, the effective electron-phonon coupling
is very strong owing to the confinement of their low dimensional
nature. With this feature, many intriguing phenomena related to
polarons and bipolarons can occur in these materials. Electrons in a
polar or ionic crystal are influenced by the phonon field giving
rise to polaronic effects \cite{DEVR, WS}. When the attractive
interaction of the electron-phonon coupling is strong enough to
overcome the Coulomb repulsion between two electrons, the two
electrons plus their associated phonon field can form a bound state.
Such a composite quasi-particle is referred to as a bipolaron. If
the bipolaron is extended over many lattice cells, it is called {\it
large} bipolaron. In contrast, if the bipolaron is localized at one
lattice site, it is called {\it small} bipolaron \cite{ED}.

Bipolarons have been extensively discussed both for fundamental
theoretical reasons and for their importance in semiconductor
materials. With recent advances in the creation of nanocrystals and
semiconductor nanostructures, strong electron-phonon coupling is
realized due to quantum confinement effects ~\cite{LA, HRR, Q, P,
SR, MDV, QH}, and the properties of bipolarons in low dimensional
systems are of growing interest. This problem is also relevant to
the proposal of the bipolaronic mechanism for electron pairing in
the $CuO_2$ plane in high $T_c$ cuprates \cite{YHK, ED}. Recently,
Schoenes and co-workers \cite{SCH} reported that some
superconductors such as $YBa_2Cu_4O_8$ have strong electron-phonon
coupling, which also boosts the theoretical research on bipolarons.

Though the bipolaron ground state has been extensively discussed in
the past decade \cite{FBA, VGS, CHQ, LFB, MSC, PEPF, SRT, RYH2}, the
excited states have, so far, been studied much less. At the same
time, knowledge of the excited states of the bipolaron is related to
electron transport, photoluminescence and photoemission \cite{AM,
DHPARK, YHK}. Several works have been devoted to excited states of
Fr\"{o}hlich bipolarons \cite{HUY, VPP, Sahoo, KAN, VKM}. Huybrechts
\cite{HUY} developed a Lee-Low-Pines variational method and studied
the ground and excited states of a single optical polaron. Smondyrev
{\it et al.} \cite{VPP} calculated the energy spectra of the
one-dimensional bipolaron in the strong-coupling limit. More
recently, Sahoo \cite{Sahoo} developed the Landau-Pekar variational
method to get the ground and first excited states of the
Fr\"{o}hlich bipolaron in a multidimensional ionic crystal in the
strong-coupling limit.

In this paper we adopt the Fr\"{o}hlich Hamiltonian for the large bipolaron
which consists of the kinetic energies of two electrons, their interactions
with the phonon field, and the screened Coulomb repulsion between them. We
extend the Huybrechts variational approach (LLP-H) to the analysis of the
bipolaron ground state in Sec. II. For comparison, different wavefunctions
are used to calculate the ground-state energy of the bipolaron in Sec. III,
and the best wavefunction of the relative motion is obtained. In Sec. IV,
after averaging over the relative motion, the energies of the first relaxed
excited state and Franck-Condon excited state of the bipolaron in two and
three dimensions are obtained for the whole coupling parameter range, and
discussions of these two types of excited states are given. Finally,
comparison of our results with previously published data is given in Sec. V.
Our conclusions are presented in Sec. VI.

\section{Calculation of the ground-state energy}\label{GS energy}

The Fr\"{o}hlich Hamiltonian for the polaron in $N$ dimensions
(ND) has been derived by Peeters {\it et al} \cite{PEE}.
Accordingly, the Hamiltonian describing a system of two electrons
interacting with a longitudinal optical ($LO$) phonon field may be
written as
\begin{eqnarray}
H=&&\sum_{i=1,2}\left[ \frac{{\bf p}_i^2}{2m}+\sum_{{\bf k}}(V_{{\bf k}}a_{%
{\bf k}}e^{i{\bf k}\cdot {\bf r}_i}+h.c.)\right]
\nonumber\\
&&+\sum_{{\bf k}}\hbar \omega _{{\bf k}}a_{{\bf k}}^{+}a_{{\bf
k}}+U(\left| {\bf r}_1-{\bf r}_2\right| ),
\end{eqnarray}
where all vectors are $N$-dimensional ($N=2,3$), ${\bf r}_i$ (${\bf p}_i$)
is the position (momentum) operator of the $i$th electron ($i=1,2$) and $m$
is the effective electron band mass in the parabolic approximation. $a_{{\bf %
k}}^{+}$ and $a_{{\bf k}}$ are respectively the creation and annihilation
operators of the $LO$ phonons with the wave vector ${\bf k}$. Here we should
mention that the impurity-phonon interactions have already been eliminated
so that we assume $\omega _{{\bf k}}=\omega _{LO}$. For convenience in the
following, units are taken where $m=\hbar =\omega _{LO}=1$, and the
Hamiltonian for the ND Fr\"{o}hlich bipolaron has the form:
\begin{eqnarray}
H=&&\sum_{i=1,2}\left[ \frac{{\bf p}_i^2}2+\sum_{{\bf k}}(V_{{\bf k}}a_{{\bf k}%
}e^{i{\bf k}\cdot {\bf r}_i}+h.c.)\right]\nonumber\\
&& +\sum_{{\bf k}}a_{{\bf k}}^{+}a_{%
{\bf k}}+U(\left| {\bf r}_1-{\bf r}_2\right| ).
\end{eqnarray}
In this way, energies are in units of $\hbar \omega _{LO}$ and lengths in
units of $\sqrt{\hbar /(m\omega _{LO})}$ in this article. The interaction
coefficient is
\begin{eqnarray}
V_{{\bf k}}=-i\left\{\frac{\Gamma [(N-1)/2]2^{N-3/2}\pi ^{(N-1)/2}\alpha }{%
V_Nk^{N-1}}\right\}^{1/2},
\end{eqnarray}
where $V_N$ is the ND crystal volume, $\alpha$ is the dimensionless
electron-phonon coupling constant
\begin{eqnarray}
\alpha =\frac 1{\hbar \omega _{LO}}\frac{e^2}2\left( \frac 1{\varepsilon
_\infty }-\frac 1{\varepsilon _0}\right) \left( \frac{2m\omega _{LO}}\hbar
\right) ^{1/2}\;,
\end{eqnarray}
and $\varepsilon _\infty $ $(\varepsilon _0)$ is the high-frequency (static)
dielectric constant of the medium. $U({\bf r})=U/r$ is the Coulomb
interaction potential between the two electrons where the nonscreened
electron Coulomb repulsion strength is given by $U=e^2/\varepsilon _\infty $%
, which may be rewritten as
\begin{eqnarray}
U=\frac{\sqrt{2}\alpha }{1-\eta },\qquad \eta =\frac{\varepsilon _\infty }{%
\varepsilon _0}.
\end{eqnarray}

Since the bipolaron is a composite particle, it is convenient to introduce
center-of-mass and relative coordinates and momenta, ${\bf R}=({\bf r}_1+%
{\bf r}_2)/2$, ${\bf P}={\bf p}_1+{\bf p}_2$, ${\bf r}={\bf r}_1-{\bf r}_2$,
${\bf p}=({\bf p}_1-{\bf p}_2)/2$, in which the Hamiltonian can be rewritten
as
\begin{eqnarray}
H=&&\frac{{\bf P}^2}4+2\sum_{{\bf k}}\cos \left(\frac{{\bf k\cdot r}}2%
\right)(V_{{\bf k}}a_{{\bf k}}e^{i{\bf k}\cdot {\bf
R}}+h.c.)\nonumber\\
&&+\sum_{{\bf k}%
}a_{{\bf k}}^{+}a_{{\bf k}}+{\bf p}^2+\frac Ur.
\end{eqnarray}

Up to now, an exact analytical solution for the Hamiltonian (6) is not
known. However, we can make progress by averaging (6) over the relative
wavefunction $\phi (r)$, which yields an effective Hamiltonian for the
center of mass motion,
\begin{eqnarray}
H_{eff}=&&\frac{{\bf P}^2}4+\sum_{{\bf k}}(B_{{\bf k}}a_{{\bf k}}e^{i{\bf k}%
\cdot {\bf R}}+h.c.)\nonumber\\
&&+\sum_{{\bf k}}a_{{\bf k}}^{+}a_{{\bf k}}+E_r.
\end{eqnarray}
Here $B_{{\bf k}}=2V_{{\bf k}}\langle \cos (\frac{{\bf k\cdot r}}2{\bf )}%
\rangle $, $E_r=\langle {\bf p}^2+U/r\rangle $, with $\langle \cdot \cdot
\cdot \rangle $ denoting an averaging over the wavefunction $\phi (r)$. The
new Hamiltonian (7) would be equivalent to (6) if the true wavefunction
would be known which of course is not the case. Therefore, we will use a
variational approach, two specific choices will be discussed in section III.

Note that the effective Hamiltonian (7) corresponding to the center-of-mass
motion is in essence equivalent to a single-polaron Hamiltonian. The
differences between the Hamiltonian (7) and the usual Hamiltonian of a
single polaron are the following: (i) the energy is shifted by the average
of the energy of the relative motion $E_r$, and (ii) the electron-phonon
interaction coefficient $B_{{\bf k}}$ is renormalized.

In this paper, we will follow the modified Lee-Low-Pines variational
method proposed by Huybrechts \cite{HUY} for the polaron problem.
This will allow us to obtain very accurate bipolaron energies for
all coupling regimes and calculate the ground and excited-state
energies of the Fr\"{o}hlich bipolaron. Performing the unitary
transformation $U_1$
\begin{eqnarray}
U_1=\exp \left(-ia\sum_{{\bf k}}{\bf k\cdot R}a_{{\bf k}}^{+}a_{{\bf k}%
}\right),
\end{eqnarray}
the Hamiltonian (7) can be transformed into
\begin{eqnarray}
H_{eff}=&&\frac 14\left({\bf P-}a\sum_{{\bf k}}{\bf k}a_{{\bf k}}^{+}a_{{\bf k}%
}\right)^2\nonumber\\
&&+\sum_{{\bf k}}\{B_{{\bf k}}^{\star }a_{{\bf k}}^{+}\exp [-i(1-a)%
{\bf k\cdot R]}+h.c.\}\nonumber\\
&&+\sum_{{\bf k}}a_{{\bf k}}^{+}a_{{\bf k}}+E_r,
\end{eqnarray}
where $a$ is a parameter. In the limit $a\rightarrow 0$, the present
calculation is identical to the strong-coupling regime, whereas the case $%
a\rightarrow1$ corresponds to weak electron-phonon interaction.

Following Huybrechts, we introduce the creation and annihilation operators $%
b_j^{+}$ and $b_j$ for electrons
\begin{eqnarray}
P_j &=&(\lambda )^{1/2}(b_j^{+}+b_j),  \nonumber \\
R_j &=&i\left( \frac 1{4\lambda }\right) ^{1/2}(b_j-b_j^{+}),
\end{eqnarray}
where the index $j$ refers to the $N$ space directions and $\lambda $ is a
variational parameter. Rewriting the Hamiltonian (9) using (10) one gets
\begin{eqnarray}
H_{eff}^{^{\prime }}=H_1^{^{\prime }}+H_2^{^{\prime }}+E_r,
\end{eqnarray}
\begin{eqnarray}
H_1^{^{\prime }} &=&\frac \lambda 2\left( \sum_jb_j^{+}b_j+\frac N2\right)
+\sum_{{\bf k}}\left( 1+\frac{a^2k^2}4\right) a_{{\bf k}}^{+}a_{{\bf k}}
\nonumber \\
&&+\sum_{{\bf k}}\bigg\{{B_{{\bf k}}^{\star }a_{{\bf k}}^{+}\exp }\left[ {-%
\frac{(1-a)^2k^2}{8\lambda }}\right]\nonumber \\
 &&\times{\exp }\left[ {-(1-a)\sqrt{\frac 1{%
4\lambda }}\sum_jk_jb_j^{+}}\right]  \nonumber \\
&&\times {\exp }\left[ {(1-a)\sqrt{\frac 1{4\lambda }}\sum_jk_jb_j}\right] {%
+h.c.\bigg\}}\nonumber \\
&&+\frac \lambda 4\sum_j(b_j^{+}b_j^{+}+b_jb_j),
\end{eqnarray}
\begin{eqnarray}
H_2^{^{\prime }}&=&\frac{a^2}4\sum_{{\bf k,k}^{^{\prime }}}{\bf k}\cdot {\bf k}%
^{^{\prime }}a_{{\bf k}}^{+}a_{{\bf k}^{^{\prime }}}^{+}a_{{\bf k}}a_{{\bf k}%
^{^{\prime }}}\nonumber \\
&&-\frac{a\sqrt{\lambda }}2\sum_{{\bf k}%
}\sum_jk_j(b_j^{+}+b_j)a_{{\bf k}}^{+}a_{{\bf k}}.
\end{eqnarray}

Performing now the second Lee-Low-Pines transformation:
\begin{eqnarray}
U_2=\exp \left[ \sum_{{\bf k}}(f_{{\bf k}}a_{{\bf k}}^{+}-f_{{\bf k}}^{\star
}a_{{\bf k}})\right] ,
\end{eqnarray}
one obtains
\begin{eqnarray}
H^{^{\prime \prime }} &=&\frac \lambda
2\sum_jb_j^{+}b_j+\frac{N\lambda (1-a)^2}4\nonumber\\
&&+\sum_{{\bf k}}\left( 1+\frac{a^2k^2}4\right) (a_{{\bf k}}^{+}+f_{%
{\bf k}}^{\star })(a_{{\bf k}}+f_{{\bf k}})  \nonumber \\
&&+\sum_{{\bf k}}{\ }\bigg\{{B_{{\bf k}}^{\star }(a_{{\bf k}}^{+}+f_{{\bf k}%
}^{\star })\exp }\left[ {-\frac{(1-a)^2k^2}{8\lambda }}\right]\nonumber \\
&&\times{\exp }\left[
{-(1-a)\sqrt{\frac 1{4\lambda }}\sum_jk_jb_j^{+}}\right]  \nonumber \\
&&\times {\exp }\left[ {(1-a)\sqrt{\frac 1{4\lambda }}\sum_jk_jb_j}\right] {%
+h.c.\bigg\}}\nonumber \\
&&+H_1^{^{\prime \prime }}+E_r,
\end{eqnarray}
where the part $H_1^{^{\prime \prime }}$ of the Hamiltonian (15) contains
terms of no importance for the further calculation and will be omitted. The
wavefunction describes the polaron system $\Psi (r)$ including an electron
wavefunction $\phi (r)$ and a phonon field wavefunction $\left|
f\right\rangle $, $\Psi (r)=\phi (r)\left| f\right\rangle $. For the ground
state, one has
\begin{eqnarray}
\Psi _0(r)=\phi _0(r)\left| f\right\rangle ,
\end{eqnarray}
with
\begin{eqnarray}
{b_j}\phi _0(r)=0,\text{ }\phi _0(r)=c\exp (-\lambda {\sum_jR_j^2}),
\end{eqnarray}
where $b_j$ and $\lambda $ are defined in equation (10), $\phi _0(r)$
corresponds to the ground-state wavefunction for the electron and $j$ refers
to the $j$th direction of $N$ space. For the field function in the ground
state one has
\begin{eqnarray}
\left| f\right\rangle =U_2\left| 0\right\rangle ,\text{ }a_k\left|
0\right\rangle =0,
\end{eqnarray}
where $U_2$ is defined in equation (14) and $f_k$ in $U_2$ will be obtained
by minimizing the ground-state energy of the bipolaron,
\begin{eqnarray}
\frac{\partial E_{BP}^{0,ND}}{\partial f_k^{\star }}=0.
\end{eqnarray}
Then we can directly present the bipolaron ground-state energy $%
E_{BP}^{0,ND} $ as
\begin{eqnarray}
E_{BP}^{0,ND}&=&\frac N4\lambda (1-a)^2+\sum_k\left(
1+\frac{a^2k^2}4\right) \mid f_k\mid ^2\nonumber\\
 &&+\sum_k\left\{
{{B_{{\bf k}}}^{\star }f_{{\bf k}}^{\star }\exp \left[
-\frac{(1-a)^2k^2}{8\lambda }\right] +h.c.}\right\}\nonumber\\
&&+E_r,
\end{eqnarray}
with
\begin{eqnarray}
f_k=\frac{-B_{{\bf k}}^{\star }\exp [-(1-a)^2k^2/(8\lambda )]}{1+a^2k^2/4}.
\end{eqnarray}
Using (21) in (20) we obtain the ground-state energy of the bipolaron in $N$
dimensions with an arbitrary electron-phonon coupling constant:
\begin{eqnarray}
E_{BP}^{0,ND}&=&\frac N4\lambda (1-a)^2-\sum_kB_{{\bf
k}}^2\frac{\exp [-(1-a)^2k^2/(4\lambda )]}{1+a^2k^2/4}\nonumber\\
&&+E_r.
\end{eqnarray}
The effective electron mass of ND materials which is modified by
bipolaronic effect is given by (for details see Ref.~\cite{PVA})
\begin{eqnarray}
m^{*}=2+\frac 1D\sum_{{\bf k}}({\bf k})^2\frac{B_{{\bf k}}^2\exp
[-(1-a)^2k^2/(4\lambda )]}{(1+a^2k^2/4)^3}.
\end{eqnarray}
To obtain explicit results for the ground-state energy and the effective
mass we now compute the relative wavefunction of the electron pair.

\section{Choice of the best wavefunction for relative motion}\label{relative motion}

In order to calculate the energy related to the relative motion $E_r$ and
the coefficient $B_{{\bf k}}$, we have to choose a trial wavefunction $\phi
(r)$. In the following two different functional forms will be chosen: (a)
Coulombic type
\begin{eqnarray}
\phi _n^{(c)}(r)={C}r^n\exp (-\Omega r/4),
\end{eqnarray}
and (b) oscillator type
\begin{eqnarray}
\phi _n^{(o)}(r)={C}r^n\exp (-br^2/4),
\end{eqnarray}
where $b$ and $\Omega $ are variational parameters and
$n=0,1,2,...$. Using the wavefunctions (24) or (25) to calculate
$E_r$ and $B_{{\bf k}}$, we obtain the ground-state energy of the
bipolaron from equation (22). Comparing the results obtained with
wave functions (a) and (b) allows us to determine which one leads to
the best approximation of the relative motion.

In Fig. 1(a), the ground-state energy $E_{BP}^{0,3D}$ of a 3D system
is plotted
versus $\alpha$ for different Coulombic wave functions (24) for the case $%
n=0,1,2,3$. Interestingly, the energy obtained with the function $\phi
_2^{(c)}$ ($n=2$) is smaller than those obtained by the others with $n=0,1,3$%
. From the viewpoint of the variational principle, the relative
motion is best described by the function $r^2\exp (-\Omega r/4)$
among the different Coulombic wave functions.

In Fig. 1(b), we plot $E_{BP}^{0,3D}$ (3D) as a function of $\alpha
$ for different oscillator wavefunctions (25) for $n=0,1,2$. Here,
the energy obtained with $\phi _1^{(o)}$ is lower than those by
others, indicating that $r\exp (-br^2/4)$ is superior to other
oscillator wavefunctions in describing the relative motion.
Furthermore, we find that $E_{BP}^{0,3D}$ calculated with the
oscillator wavefunction $\phi _1^{(o)}$ is smaller than that
obtained by using the best Coulombic wavefunction $\phi _2^{(c)}$.
So we conclude that the oscillator wavefunction $\phi _1^{(o)}$
reflects the relative motion best in 3D materials.

\begin{figure}
\includegraphics[width=.3\textwidth,  angle=-90]{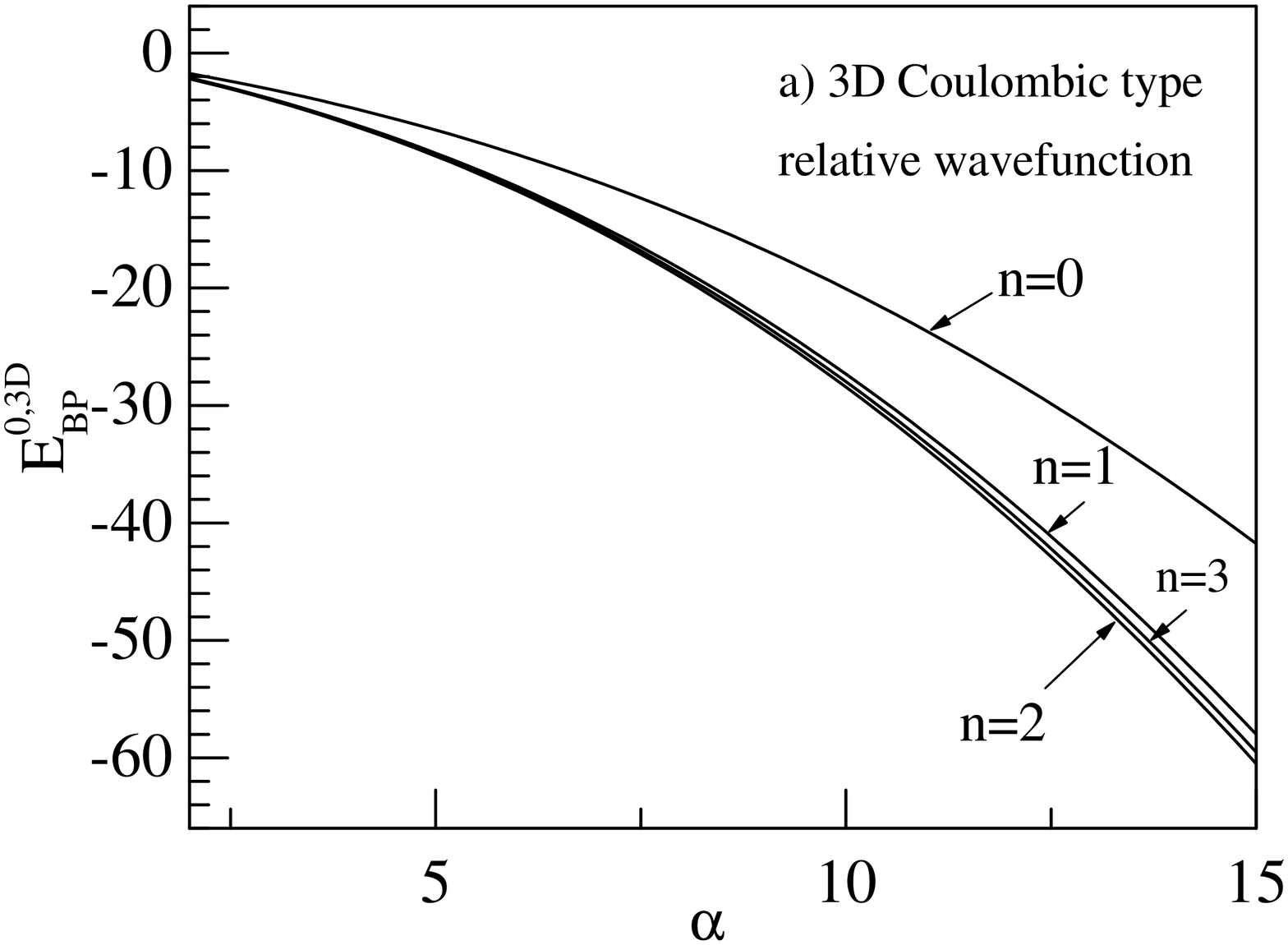}
\includegraphics[width=.3\textwidth, angle=-90]{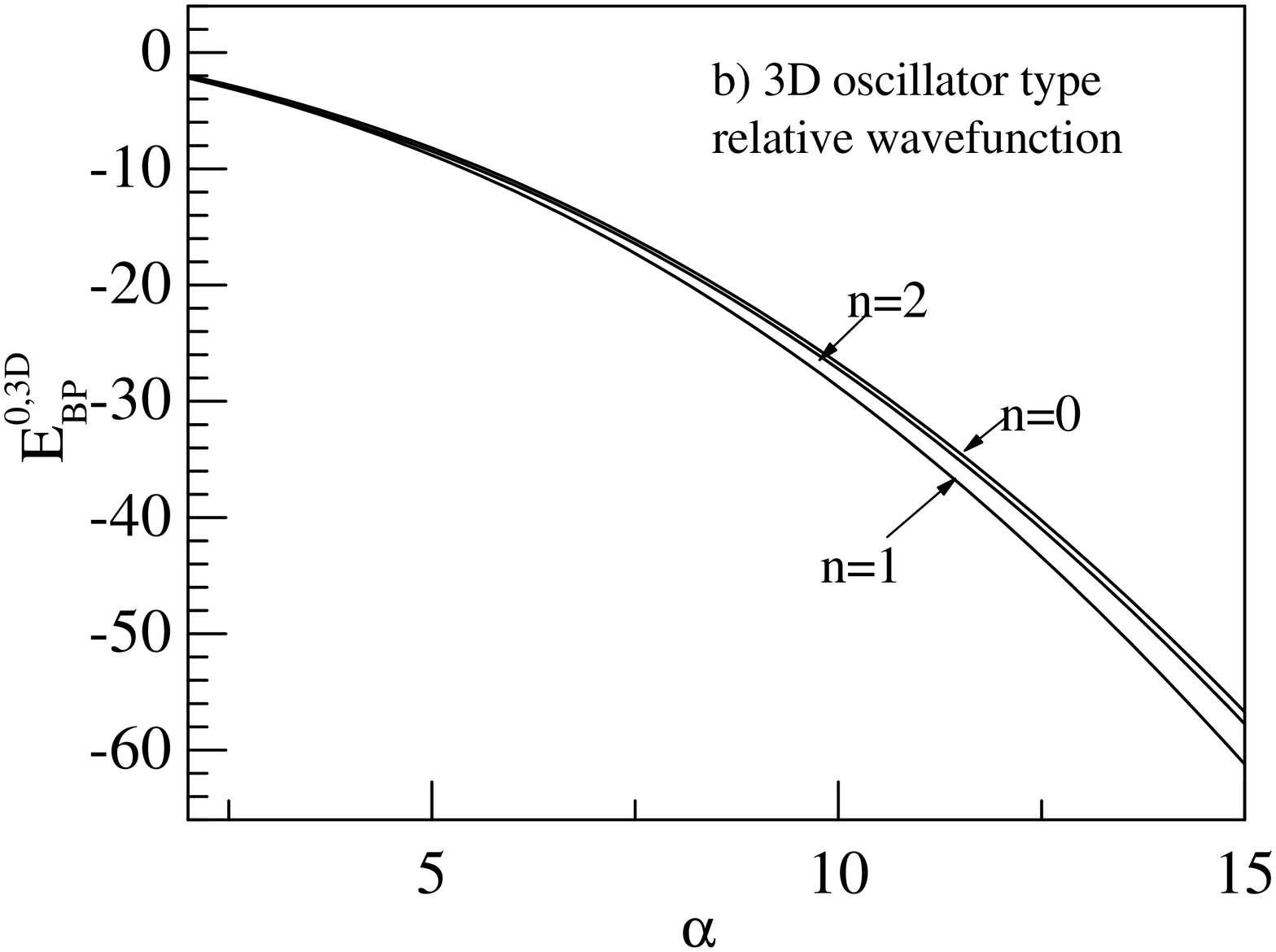}
\caption{3D bipolaron ground-state energy versus $\alpha$ using (a)
Coulombic wavefunctions (24) with $n=0,1,2,3$ for $\eta =0$; (b)
oscillator wavefunctions (25) with $n=0, 1, 2$ for $\eta =0$ (units
of energy: $\hbar \omega _{LO}$).}
\end{figure}

Next consider the 2D case. Fig. 2(a) and Fig. 2(b) present
$E_{BP}^{0,2D}$ as a function of $\alpha $ for Coulombic and
oscillator wavefunctions, respectively. In Fig. 2(a), we find that
the energy obtained by $\phi _1^{(c)}$ is lower than those by $\phi
_0^{(c)}$ and $\phi _2^{(c)}$. Fig. 2(b) shows that the energy
obtained from $\phi _1^{(o)}$ is lower than those following
from $\phi _0^{(o)}$ and $\phi _2^{(o)}$. Compared with $E_{BP}^{0,2D}$ of $%
\phi _1^{(c)}$, $E_{BP}^{0,2D}$ of $\phi _1^{(o)}$ is a little smaller,
which demonstrates that the oscillator wavefunction $\phi _1^{(o)}$ reflects
the relative motion best in 2D systems as well.

\begin{figure}
\includegraphics[width=.3\textwidth,  angle=-90]{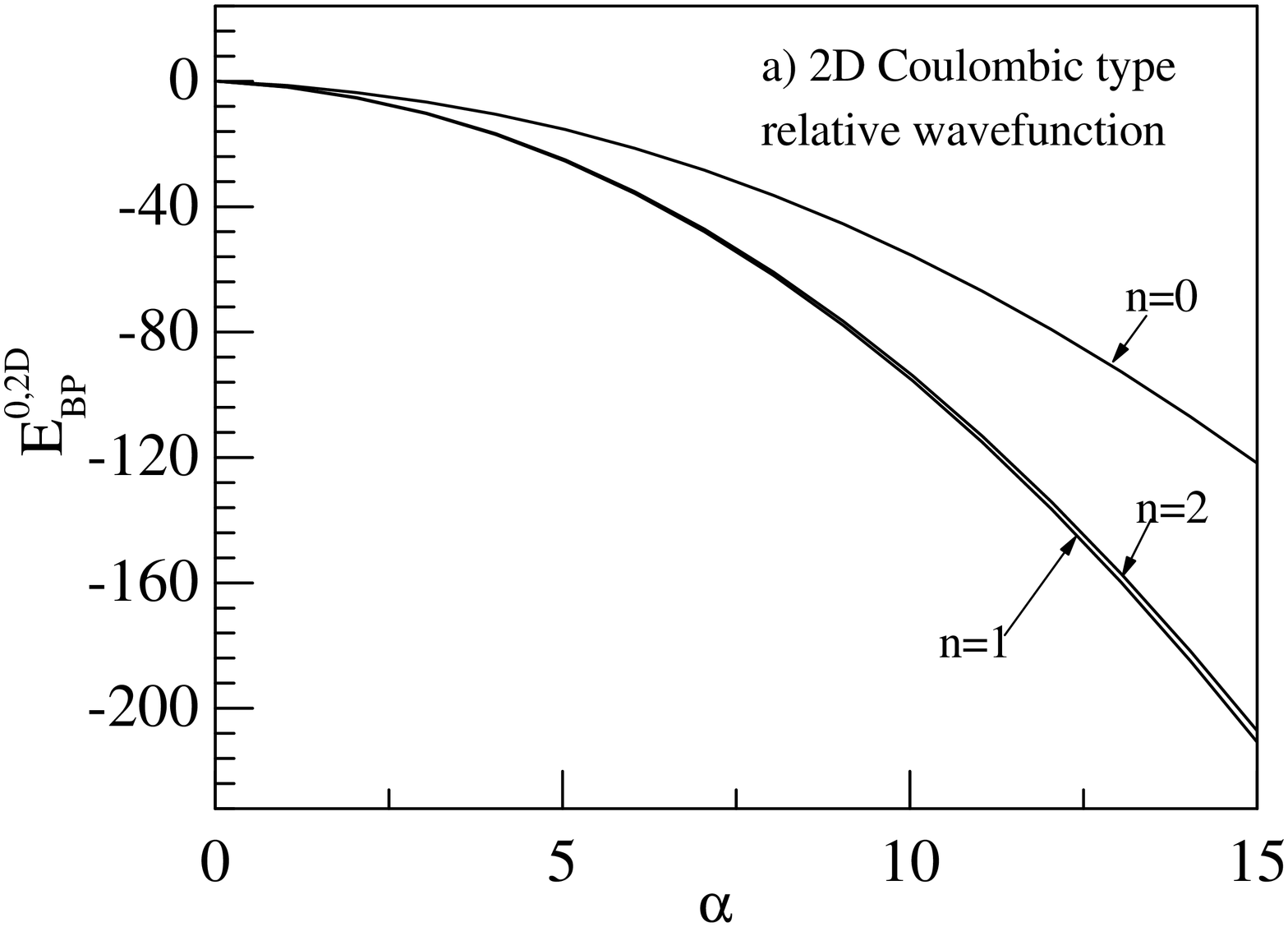}
\includegraphics[width=.3\textwidth,  angle=-90]{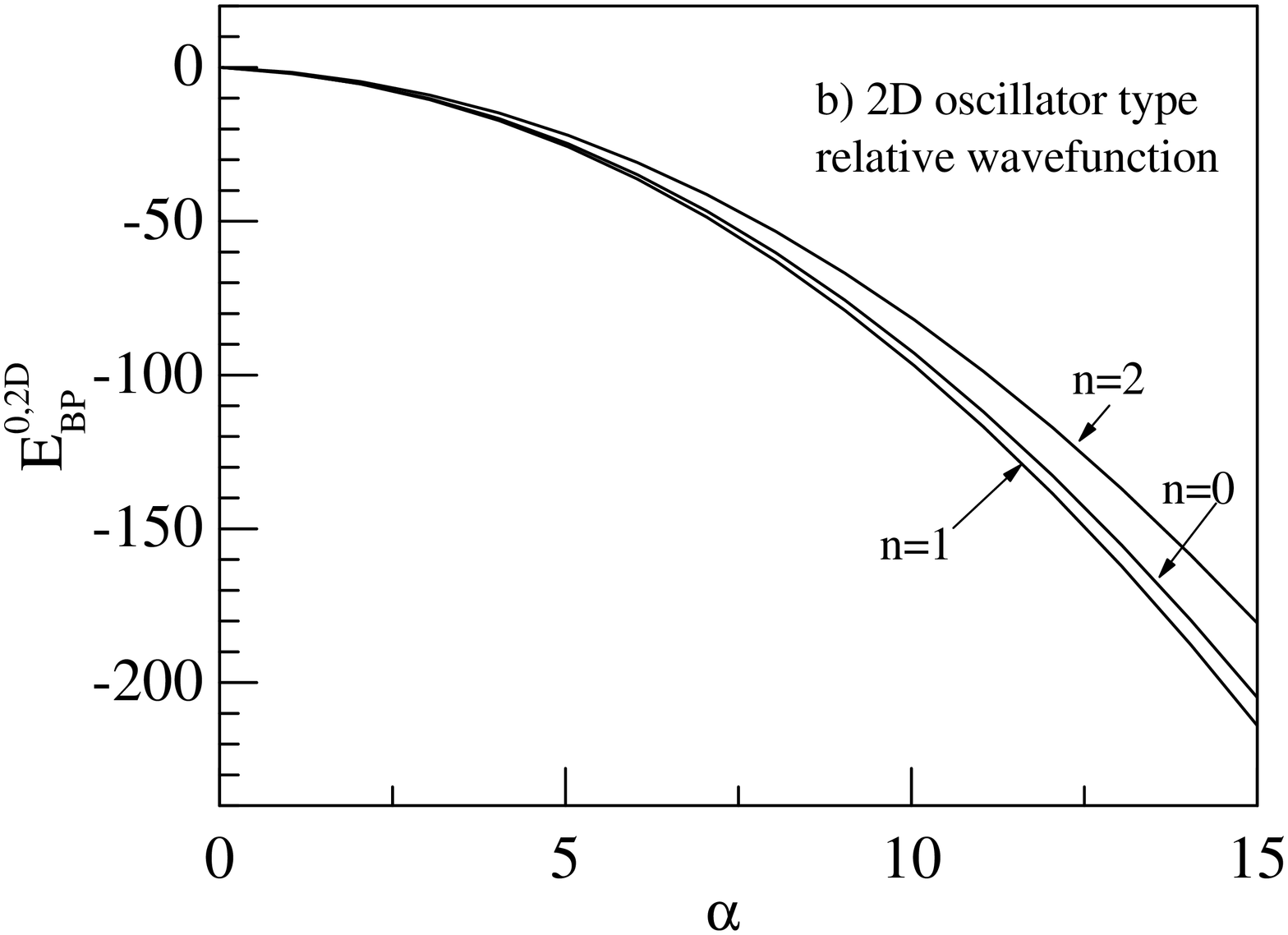}
\caption{2D bipolaron ground-state energy as a function of $\alpha$
(a) for Coulombic wavefunctions (24) with $n=0, 1, 2$ at $\eta =0$;
(b) for oscillator wavefunctions (25) with $n=0, 1, 2$ at $\eta =0$
(units of energy: $\hbar \omega _{LO}$).}
\end{figure}

Finally we calculate the energy of the relative motion $E_r$ by using the
best wavefunction $\phi _1^{(o)}$ in equation (22) and obtain the
ground-state energies of the bipolaron in 3D and 2D materials as follows
\begin{eqnarray}
E_{BP}^{0,3D}&=&\frac{3\lambda }4(1-a)^2-\frac{4\sqrt{2}\alpha }\pi
\int_0^\infty dk\left(
1-\frac{k^2}{6b}+\frac{k^4}{144b^2}\right)\nonumber\\
&&\times\frac{\exp (-gk^2)}{1+a^2k^2/4}+\frac{7b}{12}+\frac
43\sqrt{\frac b{2\pi }}U,
\end{eqnarray}
\begin{eqnarray}
E_{BP}^{0,2D}&=&\frac \lambda 2(1-a)^2-2\sqrt{2}\alpha \int_0^\infty
dk\left(
1-\frac{k^2}{4b}+\frac{k^4}{64b^2}\right)\nonumber\\
 &&\times\frac{\exp (-gk^2)}{1+a^2k^2/4}+%
\frac b2+\frac{\sqrt{2b\pi }}4U,
\end{eqnarray}
with
\begin{eqnarray}
g=\frac 1{4b}+\frac{(1-a)^2}{4\lambda },
\end{eqnarray}
and $a$, $b$ and $\lambda $ being usual variational parameters.
Comparing our bipolaron ground-state energy ($E_{BP}^{0,ND}$) with
the ground-state energy ($E_p^{0,ND}$) of a single polaron
\cite{PEE}, we find that the bipolaron is stable for $\alpha\geq
\alpha_{c}=8.1 (3.6)$ in 3D (2D). It is interesting to note that our
results are very close to the estimate $\alpha_{c}=8.1$ in 3D and
$\alpha_{c}=3.5$ in 2D in Ref.~\cite{LFB} by Luczak {\it et al.},
who obtained the bipolaron ground-state energy by a totally
different variational scheme.

Furthermore, the effective mass and root mean square separation $r_{12}$ of
two electrons can be calculated as
\begin{eqnarray}
m^{*3D}&=&2+\frac{4\sqrt{2}\alpha }{3\pi }\int_0^\infty dk\left( 1-\frac{k^2}{%
6b}+\frac{k^4}{144b^2}\right)\nonumber\\
 &&\times\frac{k^2\exp(-gk^2)}{(1+a^2k^2/4)^3},
\end{eqnarray}
\begin{eqnarray}
m^{*2D}&=&2+\sqrt{2}\alpha \int_0^\infty dk\left( 1-\frac{k^2}{4b}+\frac{k^4}{%
64b^2}\right)\nonumber\\
 &&\times\frac{k^2\exp (-gk^2)}{(1+a^2k^2/4)^3},
\end{eqnarray}
\begin{eqnarray}
r_{12}^{3D}=\left[ \frac 5b\right] ^{1/2},
\end{eqnarray}
\begin{eqnarray}
r_{12}^{2D}=\left[ \frac 4b\right] ^{1/2}.
\end{eqnarray}
The values of the variational parameters $a$, $b$ and $\lambda $
which are used in (26-32) are presented in tables 1. The effective
mass as a function of the coupling constant is shown in Fig. 3. Due
to the interaction of electron and phonon field, the effective mass
greatly increases with the coupling constant. It also can be seen
from Fig. 3 that the effective mass in 2D materials is larger than
that in 3D materials, which indicates that the electron-phonon
interaction is increased as the dimension of the material is
reduced. Similarly, in one-dimensional materials \cite{VPP}, the
bipolaron has a much larger effective mass than our results of 2D
materials.
\begin{table}[tbp]
\caption{The values of the variational parameters $\lambda$, $a$ and
$b$ which are used in the GS and FC state energies for 3D and 2D
materials.}
\begin{tabular}{lcccccc}
\hline
\hline
$\alpha$ \qquad& $\lambda^{3D}$ \qquad& $a^{3D}$ \qquad& $b^{3D}$
 \qquad& $\lambda^{2D}$ \qquad& $a^{2D}$ \qquad& $b^{2D}$ \\
\hline
1.0 & $0.628$ & $0.802$ & $0.7297$ & $1.58$ & $0.635$ & $1.795$\\
2.0 & $1.52$ & $0.638$ & $1.659$ & $4.26$ & $0.376$ & $4.743$\\
3.0 & $2.60$ & $0.507$ & $2.715$ & $8.51$ & $0.217$ & $9.396$\\
4.0 & $3.93$ & $0.398$ & $3.990$ & $14.6$ & $0.132$ & $16.11$\\
5.0 & $5.58$ & $0.310$ & $5.574$ & $22.6$ & $0.0871$ & $24.88$\\
6.0 & $7.61$ & $0.241$ & $7.530$ & $32.4$ & $0.0612$ & $35.68$\\
7.0 & $10.1$ & $0.188$ & $9.889$ & $44.1$ & $0.0452$ & $48.47$\\
8.0 & $12.9$ & $0.150$ & $12.66$ & $57.5$ & $0.0347$ & $63.25$\\
9.0 & $16.2$ & $0.121$ & $15.84$ & $72.7$ & $0.0274$ & $80.02$\\
10.0 & $19.9$ & $0.0991$ & $19.42$ & $89.8$ & $0.0222$ & $98.76$\\
11.0 & $23.96$ & $0.0826$ & $23.40$ & $108.6$ & $0.0184$ & $119.5$ \\
12.0 & $28.4$ & $0.0698$ & $27.77$ & $129.2$ & $0.0155$ & $142.2$\\
13.0 & $33.3$ & $0.0597$ & $32.53$ & $151.7$ & $0.0132$ & $166.8$\\
14.0 & $38.6$ & $0.0516$ & $37.68$ & $175.9$ & $0.0114$ & $193.5$ \\
15.0 & $44.3$ & $0.0450$ & $43.22$ & $201.9$ & $0.0099$ & $222.1$\\
 \hline
\end{tabular}
\end{table}

\begin{figure}
\includegraphics[width=.3\textwidth,  angle=-90]{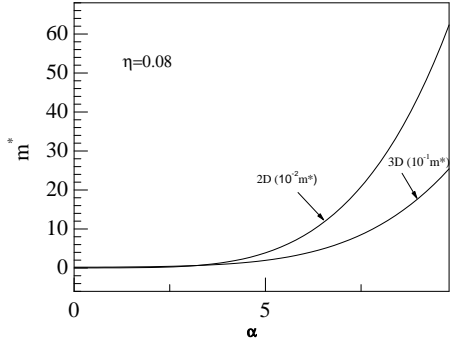}
\caption{Effective electron mass in two and three dimensions as a
function of the coupling constant at $\eta=0.08$ (units of bipolaron
effective mass: free electron mass $m _{e}$).}
\end{figure}

\section{Internal excited states}\label{ES energy}

Following the definition given by Devreese \cite{DEVR}, we compute
the relaxed excited state (RES) and Franck-Condon excited state (FC)
energies of the bipolaron. The RES is created if the electron in the
polaron is excited while the lattice readapts to the new electronic
configuration. One can imagine that the electron goes from a 1s- to
a 2p-state, while the lattice polarization in the final state is
adapted to the 2p-state of the electron. If, on the contrary, the
lattice corresponds to the electron ground state, while the electron
is excited, one speaks of a FC state.

Here we use a method of Huybrechts \cite{HUY} who computed the FC
state energy \cite{EXP} of a single polaron.
 In this section, we will develop
Huybrechts
method to calculate the RES energy of the bipolaron by adjusting $a$, $b$, $%
\lambda$ and $f_{k}$ to the relaxed excited state. Finally the FC
state and RES energies we obtain are fit for the whole
electron-phonon coupling regime.

The wavefunction for the excited polaron system is as follows
\begin{eqnarray}
\Psi _1(r)=\phi _1(r)\left| f\right\rangle ,
\end{eqnarray}
with
\begin{eqnarray}
{b_j^{+}}\phi _0(r)=\phi _1(r),\text{ }\phi _1(r)=cR_j\exp (-\lambda {%
\sum_jR_j^2}).
\end{eqnarray}
In the FC model, the electrons can be excited to a higher level in
the same potential well built up by the field of virtual phonons in
the ground state so that $\lambda $ has the same value as that in
Tab. 1. For the RES, $\lambda $ is different from that in Tab. 1 and
it can be obtained by minimizing the first excited-state energy. The
potential well should adjust to the first excited state,
Consequently, the lattice polarization in the final state is
adapted.

For the field wavefunction in the excited state one has
\begin{eqnarray}
\left| f\right\rangle =U_2\left| 0\right\rangle ,\text{ }a_k\left|
0\right\rangle =0.
\end{eqnarray}
For the FC state $f_k$ in $U_2$ is the same as that in equation (21). For
the RES $f_k$ can be obtained by minimizing the excited-state energy
according to
\begin{eqnarray}
\frac{\partial E_{BP}^{1,ND}}{\partial f_k^{\star }}=0.
\end{eqnarray}

In order to obtain the energy of the first internal excited state of the
bipolaron we calculate the average
\begin{eqnarray}
\langle 0\mid \phi _1(r)\mid H^{^{\prime \prime }}\mid \phi _1(r)\mid
0\rangle.
\end{eqnarray}
After some elementary calculus, we get the excited-state energy $%
E_{BP}^{1,ND}$ for the bipolaron in N dimensions
\begin{eqnarray}
&&E_{BP}^{1,ND}=\frac \lambda 2+\frac{N\lambda (1-a)^2}4+\sum_k\left( 1+%
\frac{a^2k^2}4\right) \mid f_k\mid ^2\nonumber\\
&&+\sum_{{\bf k}}\bigg\{ B_{{\bf k}}^{\star }f_{{\bf k}}^{\star
}\left[ 1-(1-a)^2\frac 1{4\lambda }\frac{k^2}N\right]\nonumber\\
&&\times\exp \left[ -\frac{(1-a)^2k^2}{%
8\lambda}\right]+\text{h.c.}\bigg\} +E_r,
\end{eqnarray}
with
\begin{eqnarray}
f_k^{FC}=\frac{-B_{{\bf k}}^{\star }\exp [-(1-a)^2k^2/(8\lambda )]}{%
1+a^2k^2/4},
\end{eqnarray}
for the FC state and for the RES one has
\begin{eqnarray}
f_k^{RES}&=&\frac{-B_{{\bf k}}^{\star }\exp [-(1-a)^2k^2/(8\lambda )]}{%
1+a^2k^2/4}\nonumber\\
 &&\times\left[ 1-(1-a)^2\frac 1{4\lambda
}\frac{k^2}N\right] .
\end{eqnarray}

Then we calculate the energy of the relative motion by using the best
relative wavefunction $\phi _1^{(o)}$ in equation (25) and inserting $%
f_k^{FC}$ and $f_k^{RES}$ into equation (38). Finally we obtain the FC state
and RES energies of the bipolaron in 3D and 2D materials as follows
\begin{eqnarray}
E_{BP}^{FC,3D}&&=\frac \lambda 2+\frac{3\lambda }4(1-a)^2\nonumber\\
 &&-\frac{4\sqrt{2}
\alpha }\pi \int_0^\infty dk\left(
1-\frac{k^2}{6b}+\frac{k^4}{144b^2} \right) \frac{\exp
(-gk^2)}{1+a^2k^2/4}\nonumber\\
 &&\times\left[ 1-\frac{(1-a)^2k^2}{12\lambda }
\right]+\frac{7b}{12}+\frac 43\sqrt{\frac b{2\pi }}U,
\end{eqnarray}
\begin{eqnarray}
E_{BP}^{FC,2D}&&=\frac \lambda 2+\frac \lambda 2(1-a)^2\nonumber\\
&&-2\sqrt{2}\alpha \int_0^\infty dk\left(
1-\frac{k^2}{4b}+\frac{k^4}{64b^2}\right) \frac{\exp
(-gk^2)}{1+a^2k^2/4}\nonumber\\
&&\times \left[ 1-\frac{(1-a)^2k^2}{8\lambda }\right] +\frac
b2+\frac{\sqrt{2b\pi }}4U,
\end{eqnarray}
\begin{eqnarray}
E_{BP}^{RES,3D}&&=\frac \lambda 2+\frac{3\lambda }4(1-a)^2\nonumber\\
&&-\frac{4\sqrt{2}%
\alpha }\pi \int_0^\infty dk\left( 1-\frac{k^2}{6b}+\frac{k^4}{144b^2}%
\right) \frac{\exp (-gk^2)}{1+a^2k^2/4}\nonumber\\
&&\times\left[ 1-\frac{(1-a)^2k^2}{12\lambda }%
\right] ^2+\frac{7b}{12}+\frac 43\sqrt{\frac b{2\pi }}U,
\end{eqnarray}
\begin{eqnarray}
E_{BP}^{RES,2D}&&=\frac \lambda 2+\frac \lambda 2(1-a)^2\nonumber\\
&&-2\sqrt{2}\alpha \int_0^\infty dk\left(
1-\frac{k^2}{4b}+\frac{k^4}{64b^2}\right) \frac{\exp
(-gk^2)}{1+a^2k^2/4}\nonumber \\
&&\times\left[ 1-\frac{(1-a)^2k^2}{8\lambda }\right] ^2 +\frac
b2+\frac{\sqrt{2b\pi }}4U,
\end{eqnarray}
where $E_{BP}^{1,ND}$ is expressed in units of $\hbar \omega _{LO}$.
For the FC state the values of the parameters $a$, $b$ and $\lambda
$ are taken from the calculation of the ground state energy
$E_{BP}^{0,ND}$ (Tab. 1). For the RES the values of the parameters
$\lambda $, $a$ and $b$ are obtained by minimizing the excited-state
energy $E_{BP}^{RES,ND}$ of Eq. (36) through numerical calculation,
which are presented in tables 2. From Tabs. 1 and 2, we can see that
$\lambda $ in the RES is smaller than in the ground state which
means that there is weaker potenial well built up by the phonon
field in the RES.

\begin{table}[tbp]
\caption{The values of the variational parameters $\lambda$, $a$ and
$b$ which are used in the RES energy for 3D and 2D materials.}
\begin{tabular}{lcccccc}
\hline \hline
$\alpha$ & $\lambda^{3D}$ & $a^{3D}$ & $b^{3D}$ & $\lambda^{2D}$ & $a^{2D}$ & $b^{2D}$ \\
\hline
1.0 & $1.13\times{10}{^{-9}}$ & $0.999$ & $0.631$ & $1.01\times{10}{^{-9}}$ & $0.999$ & $1.227$\\
2.0 & $3.71\times{10}{^{-9}}$ & $0.999$ & $1.170$ & $1.81\times{10}{^{-9}}$ & $0.999$ & $2.097$\\
3.0 & $2.59\times{10}{^{-9}}$ & $0.999$ & $1.556$ & $1.34$ & $0.717$ & $3.193$\\
4.0 & $0.726$ & $0.792$ & $2.085$& $3.25$ & $0.515$ & $4.671$ \\
5.0 & $1.54$ & $0.651$ & $2.685$& $5.53$ & $0.387$ & $6.478$ \\
6.0 & $2.42$ & $0.551$ & $3.337$ & $8.31$ & $0.296$ & $8.694$ \\
7.0 & $3.39$ & $0.472$ & $4.065$ & $11.6$ & $0.230$ & $11.36$\\
8.0 & $4.49$ & $0.406$ & $4.886$ & $15.5$ & $0.182$ & $14.48$\\
9.0 & $5.74$ & $0.350$ & $5.812$ & $19.9$ & $0.147$ & $18.06$ \\
10.0 & $7.13$ & $0.302$ & $6.855$ & $24.9$ & $0.121$ & $22.09$ \\
11.0 & $8.67$ & $0.262$ & $8.020$ & $30.3$ & $0.100$ & $26.57$ \\
12.0 & $10.4$ & $0.228$ & $9.312$ & $36.4$ & $0.0847$ & $31.50$\\
13.0 & $12.2$ & $0.199$ & $10.73$ & $42.9$ & $0.0724$ & $36.86$\\
14.0 & $14.3$ & $0.175$ & $12.28$ & $50.0$ & $0.0625$ & $42.67$\\
15.0 & $16.5$ & $0.155$ & $13.95$& $57.7$ & $0.0545$ & $48.91$\\
\hline
\end{tabular}
\end{table}

In Fig. 4(a) and Fig. 4(b), the ground-state energy
($E_{BP}^{0,ND}$), first RES energy ($E_{BP}^{RES,ND}$) and FC state
energy ($E_{BP}^{FC,ND}$) of the bipolaron are displayed as
functions of the electron-phonon coupling constants for 3D and 2D
materials. We find that in the whole range of electron-phonon
coupling constants, the RES energies are negative. On the
other hand, the FC state energies in 3D and 2D materials are negative when $%
\alpha >0.4$ and $\alpha >0.2$, respectively. The difference in energy, $%
\Delta E^{ND}=E_{BP}^{1,ND}-E_{BP}^{0,ND}$, yields the excitation
energy, which is related to optical absorption of bipolarons in
semiconductor materials \cite{YHK, YHKM}.

In all cases, we find that the RES energy is smaller than the FC
state energy, see Fig. 4, which is well known from absorption
spectrum calculations~\cite{DEVR, JTD, JTDE}: The absorption peak
due to a transition from the ground state to the first relaxed
excited state corresponds to the zero-phonon peak. In contrast, an
absorption transition from the GS to the FC state is accompanied
with phonon emission. If the bipolaron system is excited to the FC
state, the lattice will relax towards the RES by emission of
phonons.

\begin{figure}
\includegraphics[width=.3\textwidth,  angle=-90]{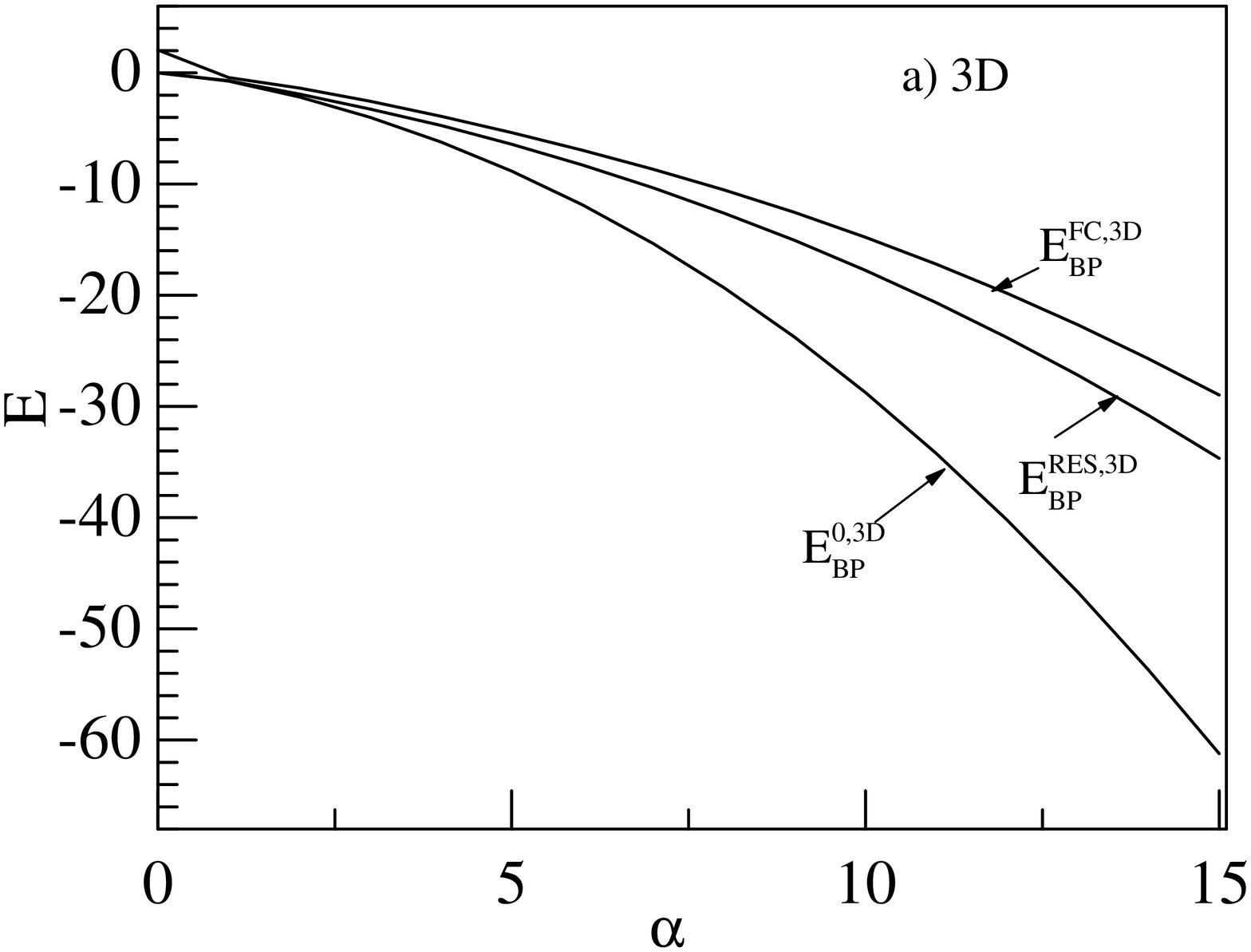}
\includegraphics[width=.3\textwidth,angle=-90]{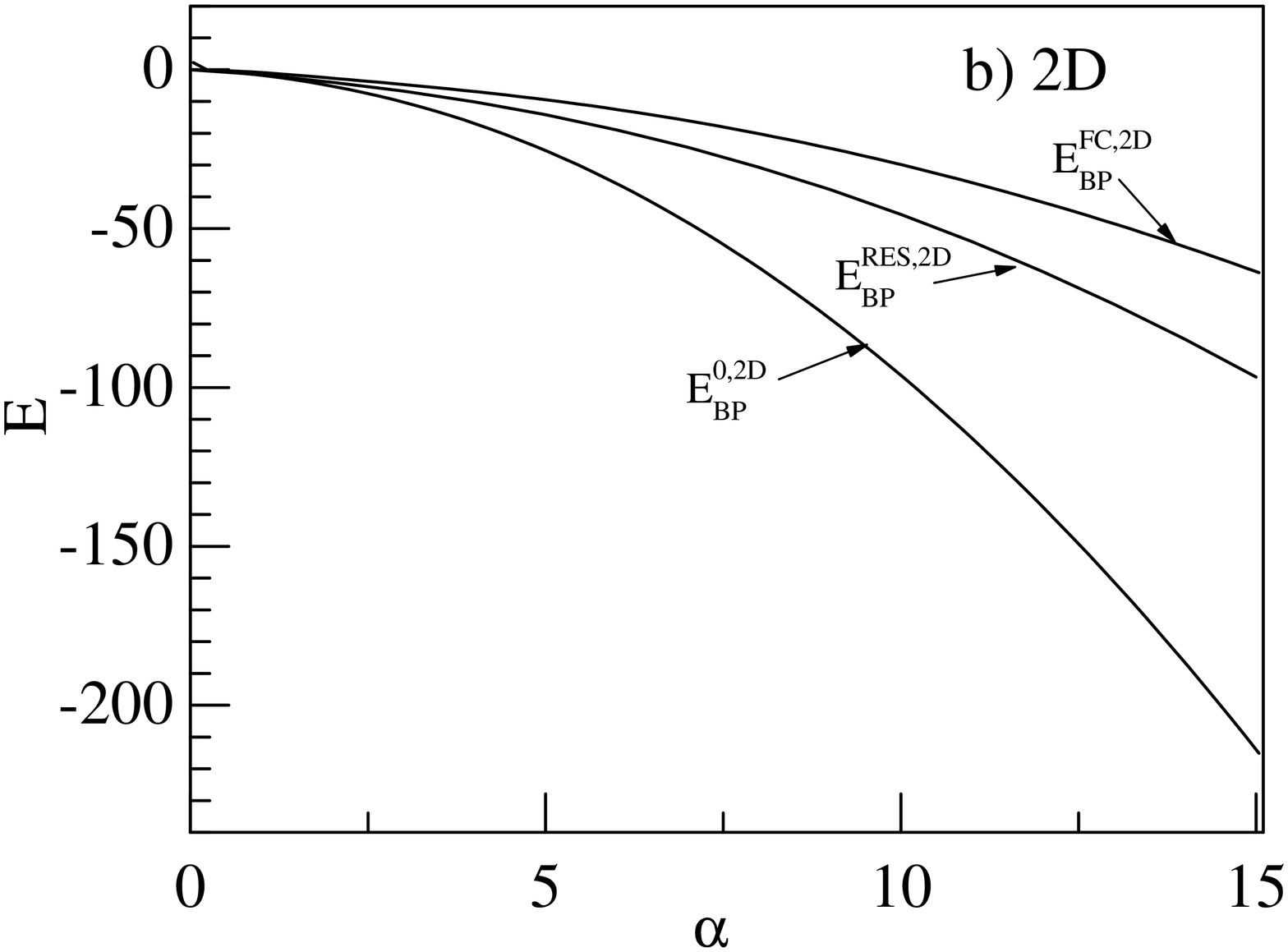}
\caption{(a) Ground-state energy ($E_{BP}^{0,3D}$), first RES energy ($%
E_{BP}^{RES,3D}$) and FC state energy ($E_{BP}^{FC,3D}$) of the
bipolaron are displayed as functions of the electron-phonon coupling
constants for 3D materials. The bipolaron is stable for $\alpha \geq
\alpha_{c}^{3D}=8.1$. (b) Ground-state energy ($E_{BP}^{0,2D}$), first RES energy ($%
E_{BP}^{RES,2D}$) and FC state energy ($E_{BP}^{FC,2D}$) of the
bipolaron are displayed as functions of the electron-phonon coupling
constants for 2D materials. The bipolaron is stable for $\alpha \geq
\alpha_{c}^{2D}=3.6$.}
\end{figure}

\section{Comparison with strong-coupling calculations}\label{comparison}

In order to assess the heuristic value of our approach, we compare
the present GS and RES energies with those obtained by Sahoo
\cite{Sahoo}, who adopted the wavefunction $\phi _0^{(o)}(r)$ and
developed a Landau-Pekar variational method to get the
ground-state and RES energies of the Fr\"{o}hlich bipolaron in the
strong-coupling limit. For the case of $N$
dimensons we obtain from Eqs. (22, 25, 38, 40) the following results when $%
\phi _0^{(o)}(r)$ is used
\begin{eqnarray}
E_{osc}^{0,ND}(0,a)&&=\frac{N\lambda }4(1-a)^2+\frac{Nb}4
+\sqrt{b/2}U\frac{\Gamma (\frac{N-1}2)}{%
\Gamma (\frac N2)}\nonumber\\
&&-\frac{2\sqrt{2}\alpha }{\sqrt{%
\pi }}\frac{\Gamma (\frac{N-1}2)}{\Gamma (\frac N2)}\int_0^\infty
\frac{\exp (-gk^2)}{1+a^2k^2/4}dk,
\end{eqnarray}
\begin{eqnarray}
E_{osc}^{RES,ND}(0,a)&=&\frac{N\lambda }4(1-a)^2
+\frac{Nb}4+\sqrt{b/2}U\frac{\Gamma (\frac{N-1}2)}{\Gamma
(\frac N2)}\nonumber \\
&&-\frac{2%
\sqrt{2}\alpha }{\sqrt{\pi }}\frac{\Gamma (\frac{N-1}2)}{\Gamma (\frac N2)}%
\int_0^\infty \frac{\exp (-gk^2)}{1+a^2k^2/4}\nonumber \\
&&\times[1-\frac{(1-a)^2k^2}{4N\lambda }%
]^2dk+\frac \lambda 2,
\end{eqnarray}
with $a$, $b$ and $\lambda $ being usual variational parameters,
and $g$ being defined in equation (28). If we set $a=0$ and use
the units $2m=\hbar =\omega =1$ equations (45, 46) reduce to the
results of Sahoo \cite{Sahoo}.
To perform a comparison with our results, we define the relative deviation $%
\xi =(E_{osc}^{0,ND}(0,0)-E_{BP}^{0,ND})/\left| E_{osc}^{0,ND}(0,0)\right| $%
, with $E_{osc}^{0,ND}(0,0)$ referring to the GS energies obtained by Sahoo%
\cite{Sahoo} and $E_{BP}^{0,ND}$ given by Eqs. (26, 27). Also, we define $%
\delta =(E_{osc}^{RES,ND}(0,0)-E_{BP}^{RES,ND})/\left|
E_{osc}^{RES,ND}(0,0)\right| $, $E_{osc}^{RES,ND}(0,0)$ referring to
the RES energies obtained by Sahoo \cite{Sahoo} and
$E_{BP}^{RES,ND}$ denoting our results (43, 44).

In Fig. 5, we plot $\xi $ and $\delta $ as functions of $\alpha $
for 3D and 2D materials at $\eta =0$. $\xi $ and $\delta $ are
positive in the whole coupling regime, indicating that our GS and
RES energies are smaller than those of Sahoo. It is well known that
the Landau-Pekar method works well only in the strong-coupling
regime, indeed $\xi $ and $\delta $ decrease
monotonously with the increase of $\alpha $. At large $\alpha $, $%
E_{osc}^{0,ND}(0,0)$ and $E_{BP}^{RES,ND}(0,0)$ are very close to our
results, which demonstrates the reliability of our approach. These numerical
results demonstrate that our results which are obtained by the LLP-H method
give improved GS and RES energies compared to previous strong-coupling models%
\cite{Sahoo}. Moreover, our results apply to the whole
electron-phonon coupling regime which is due to the additional
parameter $a$ in equation (8).

\begin{figure}[h]
\includegraphics[width=.3\textwidth,  angle=-90]{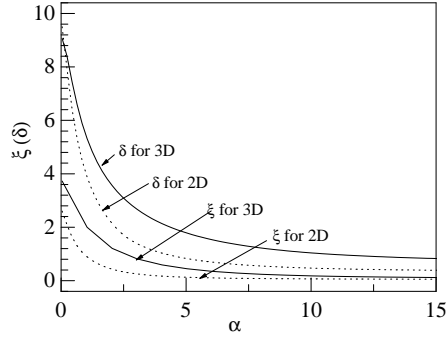}
\caption{Relative deviation of our ground-state energy and Sahoo's
result ($\xi =(E_{ osc}^{0,ND}(0,0)-E_{BP}^{0,ND})/\left| E_{
osc}^{0,ND}(0,0)\right|$) as a function of $\alpha $ for 3D and 2D
materials
at $\eta =0$; Relative deviation of our RES energy and Sahoo's result ($%
\delta =(E_{ osc}^{RES,ND}(0,0)-E_{BP}^{RES,ND})/\left| E_{
osc}^{RES,ND}(0,0)\right|$) as a function of $\alpha $ for 3D and 2D
materials at $\eta =0.$ }
\end{figure}

\section{Conclusions}\label{conclusion}

We have extended the Huybrechts variational approach (LLP-H) to the analysis
of bipolarons. By performing an average over the wavefunction of the
relative motion of the two electrons, the ground and first excited-state
energies of the bipolaron in two and three dimensions are obtained.
Numerical results show that the RES energy is smaller than the FC state
energy.

The energies we obtain are applicable to arbitrary values of the
electron-phonon coupling constants. Our ground-state and RES
energies are lower than the previously reported results from
Landau-Pekar method \cite {Sahoo}, which is due to the use of a
more appropriate relative wave function and the additional
parameter $a$ in LLP-H method. Based on analytical and numerical
results, we conclude that the best relative wavefunction is the
oscillator-type function $\phi _1^{(o)}$ both for the ground state
and the first excited state of the bipolaron. Our results could be
of relevance for high-$T_{c}$ superconductors where bipolarons are
expected to play an important role \cite{ED}.

\section{Acknowledgements}
Y.H. Ruan thanks Christian-Albrechts-Universit\"at zu Kiel for a
scholarship. The research was also supported by Science Foundation
of the Education Bureau of Zhejiang Province of P. R. China under
Grant No. 20040160.


\begin{thebibliography}{99}
\bibitem{EXP}
In Huybrechts' paper he talked about the RES. In fact, according to
the definition of Ref.~\cite{DEVR}, Huybrechts calculated the FC
state energy of a single polaron because he used the same parameters
$a$, $b$ and $\lambda$ in the excited state as that in the ground
state. Also, the coefficient $f_{k}$ in his calculation of the
excited state has the same value as in the ground state.
\end{thebibliography}
\end{document}